\documentclass[12pt]{article}
\usepackage{graphicx}\usepackage{amssymb}\usepackage{bm}\usepackage{amsfonts}\usepackage{amsmath}\usepackage{cases}
\newcommand{\beq}{\begin{equation}}\newcommand{\eeq}{\end{equation}}\newcommand{\beqa}{\begin{eqnarray}}
\newcommand{\eeqa}{\end{eqnarray}}\newcommand{\w}{\wedge}\newcommand{\ts}{\textstyle}
\newcommand{\nn}{\nonumber}

\newcommand{\ou}[3]{\underset{#3}{\overset{#1}{#2}}}
\newcommand{\ua}{\uparrow}
\newcommand{\da}{\downarrow}
\newcommand{\wt}{\widetilde}

\begin{document}
{\renewcommand{\thefootnote}{\fnsymbol{footnote}}

\begin{center}
{\LARGE  {d}e Sitter Spaces: Topological ramifications of gravity as a gauge theory}\\
\vspace{1.5em}
Andrew Randono\footnote{e-mail address: {\tt arandono@perimeterinstitute.ca}}
\\
\vspace{0.5em}
Institute for Gravitation and the Cosmos,\\
The Pennsylvania State
University,\\
104 Davey Lab, University Park, PA 16802, USA\\
\   \\
and \\
\   \\
The Perimeter Institute for Theoretical Physics \\
31 Caroline Street North\\
Waterloo, ON N2L 2Y5, Canada
\vspace{1.5em}
\end{center}
}

\setcounter{footnote}{0}

\begin{abstract}
We exploit an interpretation of gravity as the symmetry broken phase of a de Sitter gauge theory to construct new solutions to the first order field equations. The new solutions are constructed by performing large $Spin(4,1)$ gauge transformations on the ordinary de Sitter solution and extracting first the tetrad, then the induced metric. The class of metrics so obtained is an infinite class labelled by an integer, $q$. Each solution satisfies the local field equations defining constant positive curvature, and is therefore locally isometric to de Sitter space wherever the metric is non-degenerate. The degeneracy structure of the tetrad and metric reflects the topological differences among the solutions with different $q$. By topological arguments we show that the solutions are physically distinct with respect to the symmetries of Einstein-Cartan theory. Ultimately, the existence of solutions of this type may be a distinguishing characteristic of gravity as a metric theory versus gravity as a gauge theory.
\end{abstract}

\section{Introduction}
Similarities between gravity and the gauge theories of the standard model abound. The geometric ingredients of general relativity in the Einstein-Cartan framework include a principle G-connection, the spin connection, based on the local Lorentz group similar to connections based on the unitary groups of the standard model. In this framework extra ingredients, namely the tetrad, must be added to make contact with ordinary general relativity \cite{Utiyama:1956sy}\cite{Kibble:1961ba}. However, it has been known for some time that the tetrad and spin connection can be combined into a single connection based on the Poincar\'{e}, de Sitter, or anti-de Sitter group depending on the value of the cosmological constant \cite{MMoriginal}, thereby closing further the gap between gravity and an ordinary gauge theory. On the other hand, there are key differences between gravity and the gauge theories of the standard model that cannot be over-emphasized. Apart from diffeomorphism-invariance of general relativity, the most notable difference between gravity and an ordinary gauge theory lies in the local symmetries the theory retains. Whereas the gauge theories of the standard model based on principle G-connections retain full G-symmetry (linearly prior to dynamic symmetry breaking, and non-linearly after), Macdowell-Mansouri gravity based on the (A)dS group retains only the local symmetries of the Lorentz subgroup. Even at the level of the action, the exact local (A)dS symmetry is clearly broken, whereas local Lorentz symmetry is retained. In light of dynamic symmetry breaking mechanisms of the standard model, it is natural to speculate that perhaps ordinary general relativity is itself the symmetry broken phase of a more fundamental theory based on the (A)dS group as the local gauge group. In fact, semi-dynamical spontaneous symmetry breaking mechanisms have been introduced, which retain the full (A)dS symmetry but yield general relativity in the symmetry broken phase \cite{West:1978Lagrangian}\cite{Stelle:1979aj}\cite{Stelle:1979va}\cite{Fukuyama:Hamiltonian}\cite{Fukuyama:1984}\cite{Fukuyama-Ikeda}\cite{Starodubtsev:MMgravity}. However, these model remain understudied and poorly understood.

In this work we will take seriously the idea that gravity is the symmetry broken phase of a more fundamental theory and explore the consequences. We will explore generic features based on this assumption which are independent of the details of the more fundamental (A)dS theory. That is, we will explore some features of ordinary general relativity that may be unearthed by viewing the theory as a symmetry broken sector of a larger theory. More specifically, we will use familiar techniques from spontaneously symmetry broken gauge theories to derive a new class of {\it exact} solutions to the Einstein-Cartan field equations. The class of solutions is an infinite class labelled by two integers. Each solution is locally de Sitter space in any neighborhood where the metric is non-degenerate, but they differ by topological properties of the geometry. Although locally de Sitter metrics on manifolds with exotic topologies have been constructed in the past \cite{Bengtsson:1999ia}\cite{WittMorrowJones:1993}\cite{Scannell:Thesis}\cite{Witt:Designer}, these solutions have a different character in that the topology of the base manifold is fixed to be of the form $M=\mathbb{R}\times \mathbb{S}^3$. The topological differences are differences in the global structure of the geometry that is imposed on the manifold. Perhaps surprisingly, the differences in the geometries are not related to the torsional structure of the solutions -- the torsion for all the solutions we will construct is identically zero everywhere.

The symmetry breaking mechanism we will consider in this paper is {\it explicit}, as opposed to dynamic. In a dynamic scenario, the symmetry would be realized in the symmetry broken sector, albeit non-linearly. However, in the explicit scenario at hand, the symmetries of the symmetry broken sector are that of Einstein-Cartan gravity, namely\footnote{Here, and throughout, we have denoted the group of (vertical) gauge transformations of the G-bundle with base manifold $M$ by $G_M$.}, $Spin(3,1)_M \rtimes Diff_4(M)$, and the $Spin(4,1)_M$ symmetry is not realized in Einstein-Cartan theory.

In truth, the results of this paper are independent of not only the {\it details}, but even the {\it existence} of a more fundamental theory. However, this didactic stance is intended to stimulate ongoing and future research. Ultimately the goal is to find solutions or properties that would distinguish gravity as a metric theory from gravity as a gauge theory, and more generic features are better suited for this goal. 

\section{The set-up}
The basic idea is the following. Suppose there is a more fundamental theory with an exact local de Sitter symmetry (we will consider both the dS and AdS cases, but for definiteness in this section we focus on the former), whose dynamical variables consist of the de Sitter connection $A=\omega+\frac{1}{\ell}\gamma_5 \,e$ (see Appendix \ref{App} for notation and conventions), and perhaps some other dynamical fields which we need not specify for our purposes. Suppose that there exists a symmetry breaking mechanism, which we also need not specify, whereby the theory reduces to the ordinary Einstein-Cartan theory with a cosmological constant. The phase spaces of the two theories in general will be non-intersecting, but they may have some overlap for specific states shared by both theories as shown in Figure 1. We will assume only that the more fundamental theory admits solutions that are locally flat with respect to the de Sitter curvature, so that
\beq
F[A]=dA+A\w A=R[\omega]-\frac{1}{\ell^2}e\w e +\frac{1}{\ell}\gamma_5\,T=0\,.
\eeq
Since the curvature can be separated into even and odd components, this implies the two conditions
\beq
R[\omega]=\frac{1}{\ell^2}e\w e \quad\quad\quad T=D_{\omega}e=0
\eeq
all solutions of which are locally isomorphic to de Sitter space. Naturally, Einstein-Cartan theory with a positive cosmological constant also admits such solutions, so the two phase spaces share solutions of this form.

\begin{figure}
\begin{center}
\includegraphics[width=5in]{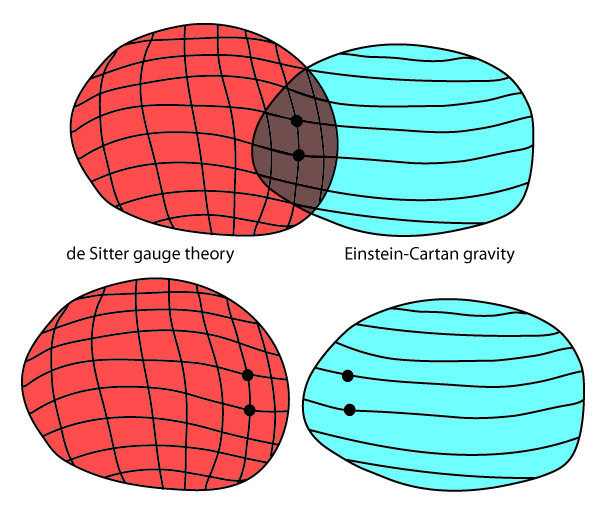}
\caption{The left side represents the phase space of the de Sitter gauge theory, and the right side represents the phase space of Einstein-Cartan gravity. The overlap region in the top diagram represents field configurations that are solutions to both theories, highlighting two points. In both cases the horizontal lines represent $Spin(3,1)_M\rtimes Diff_4(M)$ gauge orbits, and the vertical lines in the de Sitter gauge theory represent additional gauge orbits in $Spin(4,1)_M\rtimes Diff_4(M)$. Although the two solutions lie in the overlap region, to establish the two solutions as equivalent or physically distinct one must evaluate the solution spaces and their gauge symmetries separately, as shown in the bottom diagram. For the given two points, whereas the two solutions Lie on a single gauge orbit in the de Sitter gauge theory, they lie on different gauge orbits in the Einstein-Cartan theory. Thus, in the former case they are gauge equivalent solutions, whereas in the latter case they are physically distinct.}
\end{center}
\end{figure}

Now, a key property of the de Sitter gauge theory, is the existence of local de Sitter symmetry, characterized by the local gauge group $Spin(4,1)_M \rtimes Diff_4(M)$. Suppose, $A$ is some solution to the field equations of the more fundamental theory, then ${}^g A=gAg^{-1}-dg\,g^{-1}$ is also a solution for $g\in Spin(4,1)_M$. It follows that given a flat connection $A_0$ with $F[A_0]=0$, $F[{}^gA_0]=g F g^{-1}=0$. Thus both $A_0$ and ${}^g A_0$ are (gauge equivalent) solutions to the full de Sitter gauge theory. On the other hand, the phase space of Einstein-Cartan theory does not have $Spin(4,1)_M \rtimes Diff_4(M)$ as a local gauge group but only the subgroup $Spin(3,1)_M \rtimes Diff_4(M)$. Thus, {\it with respect to the symmetries of Einstein-Cartan theory}, $A$ and ${}^g A$ could, potentially, be considered physically inequivalent field configurations. This scenario is pictured in Figure 1.

The caveat is that when restricted to flat connections, gauge transformations can sometimes be identified with diffeomorphisms and vice-versa (see e.g. \cite{Carlip:Book}). This is easy to see at the infinitesimal level. Consider an infinitesimal diffeomorphism generated by a vector field $\bar{V}$ so that $A\rightarrow A'=A+\mathcal{L}_{\bar{V}}A$. From the Cartan identity for a flat connection we have
\beq
F(\bar{V})=\mathcal{L}_{\bar{V}}A-D_A (A(\bar{V})) \quad \stackrel{F[A]=0}{\longrightarrow} \quad \mathcal{L}_{\bar{V}}A=D_A(A(\bar{V}))\,.
\eeq
Identifying $\lambda=-A(\bar{V})$ as an element of the Lie algebra, an infinitesimal diffeomorphism is equivalent to an infinitesimal gauge transformation. The converse is also true provided the tetrad is invertible: an infinitesimal $Spin(4,1)_M$ gauge transformation of a flat connection is equivalent to an infinitesimal $Spin(3,1)_M \rtimes Diff_4(M)$ transformation. Thus, the identity connected component of $Spin(4,1)_M$ restricted to the space of locally flat de Sitter connections is generally related to the identity connected part of $Spin(3,1)_M \rtimes Diff_4(M)$, which {\it is} a subgroup of the symmetry group of Einstein-Cartan gravity. Thus, for an identity connected $g(x)\in Spin(4,1)_M$, even with respect to the restricted set of symmetries of Einstein-Cartan gravity, it often happens that ${}^g A$ and $A$ are gauge equivalent field configurations. On the other hand, this is not necessarily true for the {\it large gauge transformations}. It is known in $2+1$ gravity that the group of large gauge transformations bears no generic relation to the group of large diffeomorphisms, the mapping class group, of the manifold (see \cite{Carlip:Book}\cite{Matschull:1995nf} and especially \cite{Baadhio:1992jn} for a simple proof in the context of Chern-Simons theory), and it should be expected that there is no relation between the two in $3+1$ gravity as well. Although we will not present a generic proof that the two groups are distinct, we will demonstrate that the specific solutions constructed here are {\it not} related by a diffeomorphism, either large or small. 

The procedure is then the following. First we need to characterize the elements of the large gauge sector of the de Sitter group. As we will see, these elements are characterized by two ``winding numbers", $m$ and $n$ which label the homotopically inequivalent maps denoted $\underset{n}{\overset{m}{g}}\in Spin(4,1)_M$. Given a fiducial flat connection $A_0\equiv \ou{0}{A}{0}$, we then build the infinite class of flat connections
\beq
\ou{m}{A}{n}\equiv \ou{m}{g}{n} \,A \,\ou{m}{g}{n}{}^{-1}-(d\,\ou{m}{g}{n} ) \,\ou{m}{g}{n}{}^{-1} \,.
\eeq
From this we extract the tetrad and the metric
\beq
\ou{m}{e}{n}{}^I \quad \quad \quad \ou{m}{\mathfrak{g}}{n}\equiv \eta_{IJ} \,\ou{m}{e}{n}{}^I  \otimes \ou{m}{e}{n}{}^J \,.
\eeq
The advantage to extracting the metric is that it eliminates all of the gauge freedom from the local gauge group, leaving only diffeomorphism freedom. 

This procedure will yield an infinite class of metrics, all of which are locally isomorphic to de Sitter space. The task is then to determine if this class of metrics are all diffeomorphically equivalent, or if they represent distinct solutions to the Einstein-Cartan field equations, differing by some topological property. To accomplish this task, we will construct and compute an $Spin(3,1)_M\rtimes Diff_4(M)$ invariant observable that distinguishes the solutions with different values of $q\equiv m-n$.

To contrast the de Sitter group and the anti-de Sitter group, we will begin the construction using both. Eventually it will become clear that the construction gives trivial results for the anti-de Sitter group.

\section{The large gauge sectors of the de Sitter and anti-de Sitter groups}
We now wish to characterize the large gauge sectors of the de Sitter and anti-de Sitter groups. Topological solutions corresponding to the Eulidean case have been considered in the past \cite{Zanelli:NiehYan}, however, the Eulidean group, $SO(5)$ or $Spin(5)$ has very different structure than the de Sitter or anti-de Sitter groups. Some care must be taken since both of the latter groups are non-compact. 

Let us first identify the third homotopy group of both group manifolds. To do this, we will take advantage of a theorem stating that every semi-simple connected Lie group, $G$, is homeomorphic to the direct product of a maximal compact subgroup, $H$, and a (non-compact) Euclidean space (here $\approx$ means ``is homeomorphic to"):
\beq
G\approx H \times \mathbb{R}^n \quad\quad \quad dim(G)-dim(H)=n\,.
\eeq
The maximal compact subgroup is essentially unique \cite{HofmannMorris:Groups}, i.e. unique up to conjugation by elements in $G$. The key point is that the topological properties of $G$ are determined by the topological properties of $H$: since $G$ is contractible to $H$, the two spaces are homotopy equivalent. We then have
\beqa
\pi_3(G)&=& \pi_3(H \times \mathbb{R}^n) \\
&=& \pi_3(H)+ \pi_3(\mathbb{R}^n) \\
&=& \pi_3(H) +\mathbf{0}
\eeqa
since $\mathbb{R}^n$ is contractible to a point. Thus, the relevant topological properties are essentially determined by the maximal compact subgroup. 

To find the maximal compact subgroups, it is useful to first identify a basis for the Lie algebras. Using the four-dimensional Clifford algebra, (in $(-,+,+,+)$ signature), a basis for the algebras is given by
\beqa
\mathfrak{spin}(4,1)= Span\{\ts{\frac{1}{2}}\gamma^{[I}\gamma^{J]} \,,\, \ts{\frac{1}{2}}\gamma_5\gamma^K \} \quad \quad \mathfrak{spin}(3,2)=Span \{ \ts{\frac{1}{2}}\gamma^{[I}\gamma^{J]} \,,\, \ts{\frac{1}{2}}\gamma^K \}\,.
\eeqa
We can now separate out the compact generators from the non-compact generators as those elements whose one-parameter subgroups formed by exponentiation of the Lie algebra element are compact. Recalling that de Sitter space is compact in the spatial directions, the spatial pseudo-translations must form compact orbits. On the other hand, anti-de Sitter space is compact in the timelike direction (prior to taking the universal cover) and non-compact in the spacelike directions. Thus, we have
\beqa
& \mathfrak{spin}(4,1)= Span\underbrace{\{\ts{\frac{1}{2}}\gamma^{[i}\gamma^{j]}\,,\, \ts{\frac{1}{2}}\gamma_5\gamma^k \}}_{Compact} \ \oplus \ Span\underbrace{\{\ts{\frac{1}{2}}\gamma^{[i}\gamma^{0]} \,,\, \ts{\frac{1}{2}}\gamma_5\gamma^0 \}}_{Non-compact}& \\
& \mathfrak{spin}(3,2)= Span\underbrace{\{\ts{\frac{1}{2}}\gamma^{[i}\gamma^{j]} \,,\, \ts{\frac{1}{2}}\gamma^0 \}}_{Compact} \ \oplus \ Span\underbrace{\{\ts{\frac{1}{2}}\gamma^{[i}\gamma^{0]}\,,\, \ts{\frac{1}{2}}\gamma^k \}}_{Non-compact}& \,.
\eeqa
Using the Cartan-Killing metric on the Lie algebras (formed by simply taking the trace of two Lie algebra elements), it is easy to see that as a vector space the non-compact part of $\mathfrak{spin}(4,1)$ is $\mathbb{R}^4$, and the non-compact part of $\mathfrak{spin}(3,2)$ is $\mathbb{R}^6$. One can also rescale the non-compact generators by a parameter, and perform a Wigner-In\"{o}n\"{u} contraction by taking the limit as the scaling parameter goes to zero, keeping only linear terms. In this contracted limit, the non-compact generators form genuine translational subgroups $\mathbb{R}^4$ and $\mathbb{R}^6$ for the de Sitter and anti-de Sitter cases respectively. Thus, the two groups are homeomorphic to $Spin(4,1)\approx H_{dS}\times \mathbb{R}^4$ and $Spin(3,2)\approx H_{AdS}\times \mathbb{R}^6$.

We now need to determine the maximal compact subgroups $H_{dS}$ and $H_{AdS}$. In the anti-de Sitter case, it is clear that the spatial rotation generators commute with the timelike translation generators. Since the spatial rotations form the subgroup $SU(2)$, and the timelike translations form the subgroup $U(1)$, we have
\beq
Spin(3,2)\approx \left(SU(2)\times U(1)\right) \times \mathbb{R}^6 \simeq \mathbb{S}^3\times \mathbb{S}^1\times \mathbb{R}^6\,.
\eeq
Thus, we have
\beqa
\pi_3(Spin(3,2))&=& \pi_3(\mathbb{S}^3)+\pi_3(\mathbb{S}^1)+\pi_3(\mathbb{R}^6)\nn\\
&=& \mathbb{Z}+\mathbf{0}+\mathbf{0}\,.
\eeqa

We turn now to the de Sitter case. To gain some understanding of the maximal compact subgroup it is useful to work in the Dirac representation where
\beqa
\gamma^0=-i\left[\begin{array}{cc} 1 & 0 \\ 0 & -1 \end{array}\right] \quad \quad 
\gamma^i=-i\left[\begin{array}{cc} 0 & \sigma^i \\ -\sigma^i & 0 \end{array}\right] \quad \quad
\gamma_5=\left[\begin{array}{cc} 0 & 1 \\ 1 & 0 \end{array}\right] \label{DiracRep}
\eeqa
so that
\beqa
\frac{1}{2}\gamma^{[i}\gamma^{j]}=\left[\begin{array}{cc} \frac{i}{2}\epsilon^{ij}{}_k\,\sigma^k & 0 \\ 0 & \frac{i}{2}\epsilon^{ij}{}_k\,\sigma^k \end{array}\right] \quad \quad
\frac{1}{2}\gamma_5 \gamma^k=\left[\begin{array}{cc} \frac{i}{2}\sigma^k & 0 \\ 0 & -\frac{i}{2}\sigma^k \end{array}\right]\ .
\eeqa
Now define
\beqa
\tau^i_\ua &\equiv&  \frac{1}{2}\left (\frac{1}{4}\epsilon^i{}_{jk}\gamma^{[j}\gamma^{k]}+\frac{1}{2} \gamma_5 \gamma^i \right)=\left[\begin{array}{cc} \frac{i}{2}\sigma^i & 0 \\ 0 & 0 \end{array}\right] \\
\tau^i_\da &\equiv&  \frac{1}{2}\left(\frac{1}{4}\epsilon^i{}_{jk}\gamma^{[j}\gamma^{k]}-\frac{1}{2} \gamma_5 \gamma^i \right) =\left[\begin{array}{cc} 0 & 0 \\ 0 & \frac{i}{2}\sigma^i \end{array}\right]\ .
\eeqa
Since $\tau^i_\ua$ and $\tau^i_\da$ are linear combinations of the compact generators, and they clearly generate independent $SU(2)_\ua$ and $SU(2)_\da$ subgroups, the maximal compact subgroup of $Spin(4,1)$ is $Spin(4)=SU(2)\times SU(2)$. Thus, we have\footnote{More generically, the maximal compact subgroup of $Spin(p,q)$ is $(Spin(p)\times Spin(q))/\{\{1,1\},\{-1,-1\}\}$, confirming our derivation.}
\beqa
Spin(4,1)\approx Spin(4) \times \mathbb{R}^4 \simeq \mathbb{S}^3 \times \mathbb{S}^3 \times \mathbb{R}^4
\eeqa
from which we derive
\beqa
\pi_3(Spin(4,1))&=& \pi_3(\mathbb{S}^3)+\pi_3(\mathbb{S}^3)+\pi_3(\mathbb{R}^4) \nn\\
&=& \mathbb{Z}+\mathbb{Z} +\mathbf{0}\,.
\eeqa

\section{The generators of large gauge transformations}
We will now construct the generating elements of the third homotopy groups. But first, let us point out a key difference between the de Sitter and anti-de Sitter case. In the de Sitter case, the non-trivial topological properties of $Spin(4,1)$ come from the $SU(2)$ subgroups. These subgroups are generated by $\tau^i_\ua$ and $\tau^j_\da$, which are linear combinations of spatial rotations and translation generators. Thus, the group elements they generate are {\it not} contained in the $Spin(3,1)$ Lorentz subgroup. On the other hand, for the anti-de Sitter case, the non-trivial topological properties come from the $SU(2)$ subgroup generated strictly by spatial rotations, which {\it are} contained in the Lorentz subgroup. Thus, whereas in the de Sitter case there are non-trivial topological properties apart from those associated with the Lorentz subgroup, for the anti-de Sitter group all the relevant topological properties essentially come from the Lorentz subgroup itself. Since local Lorentz symmetry is an ordinary symmetry of Einstein-Cartan gravity, in the anti-de Sitter case since the resulting transformation will be an ordinary gauge transformation, albeit a large one. In particular, this means that the induced metric after the gauge transformation will be identical to the metric prior to the gauge transformation, ${}^n\mathfrak{g}=\mathfrak{g}$, since the metric is $Spin(3,1)_M$ invariant. For this reason, for the remainder of the paper we will consider only the de Sitter case.

For the de Sitter group, we can now easily construct the generators of the large gauge group using well known properties of $SU(2)$ \cite{Instantons:Examples}. Suppose the spatial hypersurface, $\Sigma$, is topologically a three-sphere, $\Sigma=\mathbb{S}^3$. Let $X^{\hat{a}}$ and $Y^{\hat{a}}$ be the embedding coordinates of $\Sigma$ as the unit three sphere in the Euclidean space $\mathbb{R}^4$, where it is understood that $\hat{a}=\{1,2,3,4\}$. Thus, $\delta_{\hat{a}\hat{b}}X^{\hat{a}}X^{\hat{b}}=1$ and $\delta_{\hat{a}\hat{b}}Y^{\hat{a}}Y^{\hat{b}}=1$. The generators of the large gauge transformations are
\beqa
g^1_\uparrow = \mathbf{1}_\downarrow+X_{\hat{4}} \,\mathbf{1}_\uparrow +X_{\hat{i}} \, 2\tau^i_{\uparrow} &=&
\left[ \begin{matrix}
X_4\,\mathbf{1}+X_{i} \,i\sigma^i & 0 \\
0 & \mathbf{1}
\end{matrix} \right]
\nn\\
g^1_\da = \mathbf{1}_\ua +Y_{\hat{4}} \,\mathbf{1}_\da +Y_{\hat{i}} \,2\tau^i_{\da}&=&
\left[\begin{matrix}
\mathbf{1} & 0 \\
0 & Y_4\,\mathbf{1}+Y_{i} \,i\sigma^i
\end{matrix}\right]
\eeqa
where
\beq
\mathbf{1}_{\ua}=
\left[ \begin{matrix} \mathbf{1} & 0 \\ 0 & 0 \end{matrix} \right]  \quad \quad  
\mathbf{1}_{\da}=\left[\begin{matrix} 0 & 0 \\ 0 & \mathbf{1} \end{matrix} 
\right] \ .
\eeq
Writing $g^n_{\ua\da}=(g^1_{\ua\da})^n$, consider the group element 
\beqa
\ou{m}{g}{n}\equiv g^m_\ua\, g^n_\da \in Spin(4,1)_M\,.
\eeqa
Consider now the change in the Chern-Simons functional for the de Sitter connection $A=\omega+\frac{1}{2\ell}\gamma_5 \gamma_I \, e^I$, under the transformation $A\rightarrow  {}^g\! A\equiv gAg^{-1} -dg\,g^{-1}$:
\beqa
Y_{CS}[{}^g\!A]&=&\frac{1}{8\pi^2}\int_{\Sigma\simeq \mathbb{S}^3}Tr\left({}^g\! A\w d{}^g\!A+\frac{2}{3}{}^g\! A\w{}^g \!A\w {}^g \!A\right) \\
&=& Y_{CS}[A] +\frac{1}{24\pi^2}\int_{\Sigma} Tr \left( dg\,g^{-1}\w dg\,g^{-1}\w dg\,g^{-1}\right)\,.
\eeqa
Writing $\ou{m}{A}{n}\equiv  \ou{m}{g}{n}\,A\,\ou{m}{g}{n}{}^{-1}- (d\ou{m}{g}{n})\,\ou{m}{g}{n}{}^{-1} $, we have
\beqa
Y_{CS}[\ou{m}{A}{n}]-Y_{CS}[A] &=& \frac{1}{24\pi^2}\int_{\Sigma} Tr \left( dg^m_\ua\,(g^m_\ua)^{-1}\w dg^m_\ua\,(g^m_\ua)^{-1}\w dg^m_\ua\,(g^m_\ua)^{-1} \right)\nn\\
& & +\frac{1}{24\pi^2}\int_{\Sigma} Tr \left( dg^n_\da\,(g^n_\da)^{-1}\w dg^n_\da\,(g^n_\da)^{-1}\w dg^n_\da\,(g^n_\da)^{-1} \right) \nn\\
&=& \frac{m}{24\pi^2}\int_{\Sigma} Tr \left( dg^1_\ua\,(g^1_\ua)^{-1}\w dg^1_\ua\,(g^1_\ua)^{-1}\w dg^1_\ua\,(g^1_\ua)^{-1} \right) \nn \\
& & +\frac{n}{24\pi^2}\int_{\Sigma} Tr \left( dg^1_\da\,(g^1_\da)^{-1}\w dg^1_\da\,(g^1_\da)^{-1}\w dg^1_\da\,(g^1_\da)^{-1} \right)\,.
\eeqa
As a manifestation of the index theorem, the integrals are related to the index of the vector fields $X^{\hat{a}}$ and $Y^{\hat{b}}$ by
\beqa
& & \frac{m}{24\pi^2}\int_{\Sigma} Tr \left( dg^1_\ua\,(g^1_\ua)^{-1}\w dg^1_\ua\,(g^1_\ua)^{-1}\w dg^1_\ua\,(g^1_\ua)^{-1} \right)\nn\\
& & \quad \quad =\frac{m}{2\pi^2}\left(\frac{1}{3!}\int_{\Sigma}\epsilon_{\hat{a}\hat{b}\hat{c}\hat{d}}\,X^{\hat{a}}\,dX^{\hat{b}} \w dX^{\hat{c}} \w dX^{\hat{d}}\label{IndexX}\right)\\
& & 
\frac{n}{24\pi^2}\int_{\Sigma} Tr \left( dg^1_\da\,(g^1_\da)^{-1}\w dg^1_\da\,(g^1_\da)^{-1}\w dg^1_\da\,(g^1_\da)^{-1} \right) \nn\\
& &\quad \quad = \frac{n}{2\pi^2}\left(\frac{1}{3!}\int_{\Sigma}\epsilon_{\hat{a}\hat{b}\hat{c}\hat{d}}\,Y^{\hat{a}}\,dY^{\hat{b}} \w dY^{\hat{c}} \w dY^{\hat{d}}\right) \label{IndexY}\,.
\eeqa
The integrals in parentheses in (\ref{IndexX}) and (\ref{IndexY}) are identified with $Vol(\mathbb{S}^3)=2\pi^2$, yielding the final result
\beq
Y_{CS}[\ou{m}{A}{n}]-Y_{CS}[A] =m+n\,.
\eeq
We conclude that the group elements $g^1_\ua\!: \mathbb{S}^3 \rightarrow Spin(4,1)$ and $g^1_{\da}\!:\mathbb{S}^3 \rightarrow Spin(4,1)$ are the generators of $\pi_3(Spin(4,1))=\mathbb{Z}+\mathbb{Z}$.

\section{Finding the induced metric}
We will now proceed to find the tetrad and metric induced by the large de Sitter transformation. The construction proceeds as follows. Begin with a fiducial representation of de Sitter space defined by a tetrad $e=\frac{1}{2}\gamma_I\,e^I$, and its torsion free spin-connection $\omega=\frac{1}{4}\gamma_I\gamma_J\,\omega^{IJ}$. Combine these two elements into a fiducial de Sitter connection $A=\omega+\frac{1}{\ell}\gamma_5\,e$, where $\ell=\sqrt{\frac{3}{\Lambda}}$ and $\Lambda$ is the cosmological constant. Call this fiducial de Sitter connection $\ou{0}{A}{0}$. We then explicitly construct the gauge transformed field $\ou{m}{A}{n}$ written in terms of the fiducial tetrad and spin connection, and the vector fields $X^{\hat{a}}$ and $Y^{\hat{b}}$. Once we have this connection, we proceed to extract the tetrad identified as the odd component of the connection given by the formula:
\beq
\ou{m}{e}{n}=\frac{\ell}{2}\gamma_5 \left(\ou{m}{A}{n}-\gamma_5 \ou{m}{A}{n} \gamma_5\right)= \frac{\ell}{2}[\gamma_5 \,,\, \ou{m}{A}{n}]\,.
\eeq
The metric can then be induced by
\beq
\ou{m}{\mathfrak{g}}{n} \equiv Tr(\ou{m}{e}{n} \otimes \ou{m}{e}{n})=\eta_{IJ}\,\ou{m}{e}{n}{}^I \otimes \ou{m}{e}{n}{}^J\,.
\eeq

Let us first set up some preliminaries. The calculations will be drastically simplified by first obtaining a convenient form for the group element $\ou{m}{g}{n}$. Since the vector field $X^{\hat{a}}$ has unit norm under the Euclidean inner product at each point of the manifold, it defines a point on the three-sphere embedded in four dimensional Euclidean space. Thus, it is convenient to express the vector field in terms of the polar angles of the three sphere. Thus, define
\beqa
X^{\hat{1}}&=& \sin{\chi}\sin{\theta}\cos{\phi} \nn\\
X^{\hat{2}} &=& \sin{\chi}\sin{\theta}\sin{\phi} \nn\\
X^{\hat{3}} &=& \sin{\chi}\cos{\theta} \nn\\
X^{\hat{4}}&=& \cos{\chi}\,.
\eeqa
Identifying the coordinates $\{\chi, \theta, \phi\}$ as the coordinates of the spatial three sphere of the  de Sitter spacetime manifold $\Sigma \simeq \mathbb{S}^3$, then gives the map $\Sigma\simeq \mathbb{S}^3 \rightarrow SU(2)_{\ua}\simeq \mathbb{S}^3$. Similarly for $Y^{\hat{a}}$--the only subtlety is that the two vector fields do not necessarily have to be defined with respect to the same origin on the spatial three sphere, thus we write $Y^{\hat{a}}$ in terms of an alternative set of polar angles $\{\chi',\theta',\phi'\}$. With these identifications, it can be shown that the group elements $g^m_{\ua}$ and $g^n_{\da}$ become
\beqa
g^m_{\ua} &=& \left[\begin{matrix} \cos{m\chi}\,\mathbf{1} + \sin{m\chi}\,\widetilde{X}_{i}\,i\sigma^i & 0 \\
0 & \mathbf{1} 
\end{matrix}\right] \\
g^n_{\da} &=& \left[\begin{matrix}  \mathbf{1} & 0 \\
0 & \cos{n\chi'}\,\mathbf{1} + \sin{n\chi'}\,\widetilde{Y}_{i}\,i\sigma^i 
\end{matrix}\right]
\eeqa
where $\widetilde{X}^{i}\equiv X^{\hat{i}}/\sin{\chi}$ so that $\wt{X}_{i}\wt{X}^{i}=1$ defines a 2-sphere (recall $\{i,j,k\}$ range from one to three), and similarly for $\wt{Y}^{i}$. 

Working in the Dirac representation for the gamma-matrices given above (\ref{DiracRep}), the connection can be written:
\beq
A=\left[ \begin{matrix} W^i_+\,\tau_i & -i({\omega^i}_0\,\tau_i+\frac{1}{2\ell} e^0) \\
- i({\omega^i}_0\,\tau_i-\frac{1}{2\ell} e^0) & W^i_-\,\tau_i 
\end{matrix}\right]
\eeq
where $W^i_\pm \equiv \frac{1}{2}{\epsilon^i}_{jk}\,\omega^{jk} \pm \frac{1}{\ell}e^i=\omega^i\pm  \frac{1}{\ell}e^i$ and $\tau^i=\frac{i\sigma^i}{2}$. 

With these preliminaries, the remainder of the calculation reduces to lengthy but straight forward matrix algebra, so we will present only the end result. Without simplifying assumptions, the general solution is somewhat complicated. For the time component of the tetrad the general result is
\beqa
\ou{m}{e}{n}{}^0 &=& \left(\cos{m\chi}\cos{n\chi'}+\sin{m\chi}\sin{n\chi'} \,\wt{X}_i\wt{Y}^i \right)\,e^0 \nn\\
& & -\ell \left( \sin{m\chi}\cos{n\chi'}\,\wt{X}_i\,{\omega^i}_0-\cos{m\chi}\sin{n\chi'} \,\wt{Y}_i\,{\omega^i}_0 \right) \nn\\
& & +\ell \sin{m\chi}\sin{n\chi'}\,\epsilon_{ijk} \wt{X}^i\wt{Y}^j {\omega^k}_0 \,
\eeqa
and for the spatial components we have
\beqa
\ou{m}{e}{n}{}^i &=& \frac{\ell}{2}\left( \cos{2m\chi}\,W^i_+ -\sin{2m\chi} \,{\epsilon^i}_{jk}\,\wt{X}^j\,W^k_+ +2\sin^2{m\chi}\,\wt{X}_k\,W^k_+\,\wt{X}^i \right) \nn\\
& & - \frac{\ell}{2} \left( 2m\,\cos{2m\chi}\,\wt{X}^i\,d\chi +\sin{2m\chi}\,d\wt{X}^i 
-2\sin^2{m\chi}\,{\epsilon^i}_{jk}\wt{X}^j\,d\wt{X}^k \right) \nn\\
& & -\{ m\rightarrow n, \wt{X}^i\rightarrow \wt{Y}^i , \chi\rightarrow \chi' , W^i_+\rightarrow W^i_-\}\,.
\eeqa
The time components simplify when it is assumed that $\wt{X}^i=\wt{Y}^i$ and $\chi=\chi'$, and we will assume this for the rest of the paper. Defining $p\equiv m+n$ and $q\equiv m-n$, in this case we have
\beq
\ou{m}{e}{n}{}^0 =\cos{q\chi}\,e^0-\ell\sin{q\chi}\,\wt{X}_i {\omega^i}_0\,.
\eeq
and
\beqa
\ou{m}{e}{n}{}^i &=& \cos{p \chi} \cos{q\chi}\,e^i-\sin{p\chi}\cos{q\chi} \,{\epsilon^i}_{jk}\wt{X}^j\,e^k \nn\\
& &+(1-\cos{p\chi}\cos{q\chi})\,\wt{X}^i\,\wt{X}_j \,e^j \nn\\
& &- \ell \,\cos{p\chi}\sin{q\chi}\,{}^3\!D\wt{X}^i \nn\\
& & + \ell \sin{p\chi}\sin{q\chi}\,{\epsilon^i}_{jk}\wt{X}^j\,{}^3\!D\wt{X}^k \nn\\
& & -\ell\,q\, \wt{X}^i\,d\chi \,,
\eeqa
where we have defined ${}^3\! D\wt{X}^i=d\wt{X}^i+{\omega^i}_j\,\wt{X}^j$. The corresponding spin connection can also be computed by calculating $\ou{m}{\omega}{n}=\frac{1}{2}(\ou{m}{A}{n}+\gamma_5\ou{m}{A}{n}\gamma_5)$, yielding (recall $\omega^i \equiv \frac{1}{2}{\epsilon^i}_{jk}\,\omega^{jk}$):
\beqa
\ou{m}{\omega}{n}{}^i{}_0 &=& \cos{p\chi}\,{\omega^i}_0+\frac{1}{\ell}\sin{q\chi}\,\wt{X}^i\,e^0 \nn\\
& & -\sin{p\chi}\,{\epsilon^i}_{jk}\,\wt{X}^j{\omega^k}_0   +(\cos{q\chi}-\cos{p\chi})\,\wt{X}^i\,\wt{X}_k{\omega^k}_0
\eeqa
and
\beqa
\ou{m}{\omega}{n}{}^i &=& \omega^i  +(1-\cos{p\chi}\cos{q\chi})\,{\epsilon^i}_{jk}\,\wt{X}^j \,{}^3D\wt{X}^k \nn\\
& & -\frac{1}{\ell}\,\sin{p \chi}\sin{q\chi}\,e^i \nn\\
& & +\frac{1}{\ell}\,\sin{p\chi}\sin{q\chi} \,\wt{X}^i\,\wt{X}_ke^k   \nn\\
& & -\frac{1}{\ell}\,\cos{p\chi}\sin{q\chi} \,{\epsilon^i}_{jk}\wt{X}^j e^k\nn\\
& & -\sin{p\chi}\cos{q\chi}\,{}^3D\wt{X}^i  \nn\\
& & -p\,\wt{X}^i\,d\chi \,.
\eeqa

For writing the metric, $\ou{m}{\mathfrak{g}}{n}=\eta_{IJ} \ou{m}{e}{n}{}^I \otimes \ou{m}{e}{n}{}^J$ , it is convenient to define the four vector $\widetilde{X}^I=(0,\wt{X}^i)$ so that the covariant derivative becomes $D_{\omega}\wt{X}^i={}^3\! DX^i$ and $D_{\omega}\wt{X}^0={\omega^{0}}_{i}\,\wt{X}^i$. Written in terms of the fiducial tetrad and spin connection, the induced metric is given by
\beqa
\ou{m}{\mathfrak{g}}{n}&=&\ \  \cos^{2}{q\chi}\,\mathfrak{g} \nn\\
& & +\sin^2{q\chi}\,\left(\wt{X}_I\,e^I \otimes \wt{X}_J\,e^J \right)\nn \\
& & +\ell^2\,\sin^2{q\chi}\,\left( D_{\omega} \wt{X}_I \otimes D_\omega \wt{X}^I \right)\nn \\
& & +\ell^2\,q^2\,\left(d\chi \otimes d\chi \right) \nn\\
& & -\ell\,\sin{q\chi}\cos{q\chi}\,\left(e^I\otimes D_{\omega}\wt{X}_I +  D_{\omega}\wt{X}_I\otimes e^I \right) \nn\\
& & -\ell\,q\,\left( d\chi \otimes \wt{X}_I \,e^I+\wt{X}_I \,e^I\otimes d\chi \right)\,.
\eeqa
We first note that whereas the tetrad depends on both $p $ and $q$ (equivalently, both $m$ and $n$), the metric only depends on the difference of the winding numbers: $q\equiv m-n$. Heuristically, this is because the large sector of the gauge group is formed by a combination of the three spatial rotations and the three spatial translations. On the other hand, the metric is invariant under the spatial rotations (large or small) so the induced metric only depends on one integer, $q\equiv m-n$.

We also notice that when $q=0$, so that $m=n$, we have $\ou{m}{\mathfrak{g}}{m}=\mathfrak{g}$. From the form of the tetrad, the combination of an $SU(2)_{\ua}$ transformation with winding number $m$, and a $SU(2)_{\da}$ transformation also with winding number $m$ is not trivial. However, the induced metric reveals that it is equivalent to a local Lorentz transformation. On closer inspection, the corresponding group element can be written
\beq
\ou{m}{g}{m}=g^m_{\ua}g^m_{\da}=\cos{m\chi}\,\mathbf{1}+\sin{m\chi}\,\wt{X}_i\,{\epsilon^i}_{jk} \,{\ts\frac{1}{2}} \gamma^j \gamma^k \label{mmElement}
\eeq
revealing that it is in fact a group element in the large sector of the rotation subgroup of the the ordinary Lorentz subgroup, which leaves the metric invariant under both large and small gauge transformations.  

The task is now simply to plug in a particular fiducial tetrad and spin connection for de Sitter space. The natural form of the metric in the global $\mathbb{R}\times \mathbb{S}^3$ slicing of de Sitter space is
\beq
ds^2=-dt^2+a^2 \left(d\chi^2 +\sinh^2{\chi}(d\theta^2+\sin^2{\theta}\,d\phi^2)\right)
\eeq
with $a=\ell \cosh{t/\ell}$. We choose the diagonal tetrad
\beq
e^0=dt \quad\quad e^i=\left[\begin{array}{c}  a\,d\chi \\ a\, \sinh{\chi} \,d\theta \\ a \,\sinh{\chi} \sin{\theta} \,d\phi\end{array} \right]\,.
\eeq
With this choice, the spin connection is given by:
\beq
{\omega^i}_0=\frac{\dot{a}}{a} \,e^i=\frac{1}{\ell} \tanh(t/\ell) \,e^i 
\quad \quad {\omega^i}_j =\left[\begin{array}{ccc} 0 & -\cos{\chi}\,d\theta & -\cos{\chi}\sin{\theta}\,d\phi \\  \cos{\chi}\,d\theta & 0 & -\cos{\theta} \,d\phi \\ 
\cos{\chi}\sin{\theta}\,d\phi & \cos{\theta} \,d\phi& 0 \end{array}\right]\,.
\eeq
We notice that ${\omega^{IJ}}_t =0 $. This has the immediate consequence that $\ou{m}{\mathfrak{g}}{n}{}_{tt}=-\cos^2{q \chi} $, from which it is clear that the lapse is given by
\beq
N=\cos{q\chi} \,,
\eeq
which is zero whenever $q\chi/\pi$ is a half integer. 

\section{The determinant of the metric}
The procedure used to construct the solutions guarantees that the induced tetrad and spin-connection (written $\ou{m}{\omega}{n}$) solve the first order Einstein-field equations by virtue of satisfying the generalized constant positive curvature conditions\footnote{By construction it should be obvious that the tetrad $\ou{m}{e}{n}$ and the spin connection $\ou{m}{\omega}{n}$ satisfy the constant curvature and zero torsion conditions. However, as a check on the algebra we have confirmed this explicitly for the exact solutions given above. Incidentally, the identities hold for any {\it real} values of $m$ and $n$, but the topological interpretation of the solutions only holds for integer values.}:
\beq
F[\,\ou{m}{A}{n}\,]=0 \quad \longrightarrow \quad \begin{array}{cc} R[\,\ou{m}{\omega}{n}\,] =\frac{\Lambda}{3} \,\ou{m}{e}{n}\w \ou{m}{e}{n} \\ T[\,\ou{m}{e}{n},\ou{m}{\omega}{n}\,]=0\,. \end{array}
\eeq
However, the procedure does not ensure that the induced tetrad and metric are non-degenerate. In fact, examination of the determinant of the tetrad (which we recall is related to the determinant of the metric by $|det({e^I}_\mu)|=\sqrt{|det(\mathfrak{g}_{\mu\nu})|}$ ) reveals that the induced metric $\it{is}$ degenerate at some points on the manifold. In the gauge we have chosen, the determinant simplifies to $det({e^I}_\mu)=N det({e^i}_a)$. In addition, we have $|det({e^i}_a)|=\sqrt{|det(\mathfrak{h}_{ab})|}$ where $\mathfrak{h}_{ab}$ is the 3-metric. Thus, the determinant is
\beq
det({e^I}_\mu)=\cos{q\chi} \ det({e^i}_a),
\eeq
and the points where the determinant of the three-metric vanishes define a two-surface embedded in $\mathbb{S}^3$ where the parity of the spatial volume element reverses sign. Initial indications suggest that the surfaces defined by $det({e^i}_a)=0$ at $t=0$ are two-dimensional manifolds whose genus is equal to the absolute value of the integer $q$ (apart from the $q=0$ solution where there is no degenerate surface), though further investigation is required for verification. These features appear and disappear near the throat at $t=0$ on a time scale roughly set by $\ell$. We plot the $det({e^i}_a)=0$ surfaces at $t=0$ in Figure 1. To visualize the three-sphere, the sphere is split in half at $\chi=\pi/2$, to form two closed balls with boundaries identified. The total, integrated 3-volume of a constant-time slice reveals an interesting structure reflecting the topological differences of the solutions, which we will discuss more throughly in the next section. 

\begin{figure}
\begin{center}
\includegraphics[width=6in]{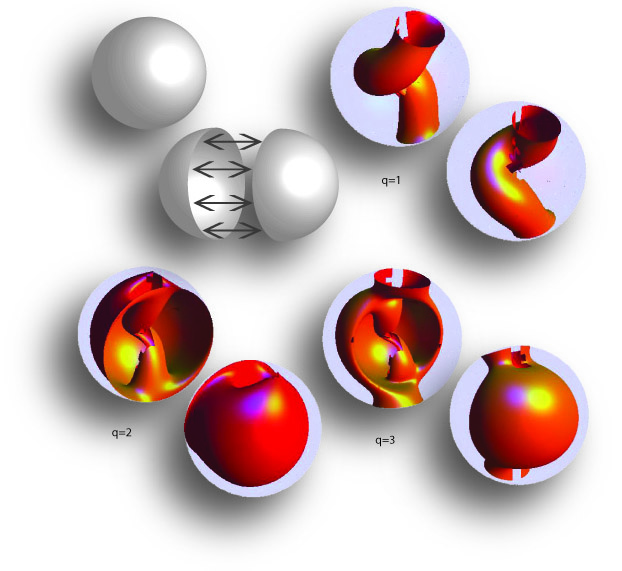}
\caption{Just as the two-sphere can be visualized by cutting the sphere into two closed disks and identifying boundary points (pictured upper left), the three-sphere can be cut into two closed balls with $\mathbb{S}^2$ boundaries identified. The surfaces defined by the vanishing of the three metric are visualized in this way for various values of $q$. The obvious defects are caused by numerical sampling errors, and are not fundamental features of the surfaces. Upper right: the $q=1$ surface clearly forms a genus-one surface, the torus $\mathbb{S}^1\times \mathbb{S}^1$, when boundary points are identified. Lower left: for $q=2$ the degenerate surface forms a genus two surface. Lower right: for $q=3$ the degenerate surface forms a genus-three surface. This leads us to conjecture that the degenerate 2-surface of the three-metric at $t=0$ for each solution solution $\{\ou{m}{e}{n},\ou{m}{\omega}{n}\}$ is a genus-$|m-n|$ surface embedded in the three-sphere.}
\end{center}
\end{figure}

\section{Are the solutions physically distinct?}
Let us now return to the question of whether the metric with non-zero $q$ are physically distinct from de Sitter space. The existence of metric degeneracies is not by itself sufficient to guarantee that the induced metrics for $q \neq 0$ are physically distinct from de Sitter space because the points where the determinant vanishes could indicate the existence of coordinate singularities that potentially could be resolved with the right diffeomorphism. To answer this question definitively, it will be sufficient to compute at least one quantity invariant under the symmetries of Einstein-Cartan theory that distinguishes ordinary de Sitter space from the other solutions. Since the solutions are related by a $Spin(4,1)_M$ transformation, we should look for a quantity that is invariant under $Spin(3,1)_M \rtimes Diff_4(M)$ but is {\it not} invariant under $Spin(4,1)_M \rtimes Diff_4(M)$. Furthermore, since the solutions are all locally de Sitter space (at points where the metric is non-degenerate), and de Sitter space is locally unique, we should look for topological quantities that may distinguish the new solutions from de Sitter space. 

Recall that there are three topological quantities defined on a manifold or region of a manifold that invariant under $Spin(3,1)_M \rtimes Diff_4(\mathcal{U})$ formed by integration of the torsion and curvature two-forms in the bulk. They are the Chern-Pontryagin class, the Nieh-Yan class, and the Euler class. Since the torsion vanishes identically everywhere, the Nieh-Yan term is identically zero. For manifolds with boundary, or non-compact manifolds where a boundary is generally added to construct well-defined expressions and later sent to infinity, the two other invariants must be modified to include boundary terms (see \cite{Eguchi:1980jx} for a discussion of these terms). These boundary terms account for the extrinsic curvature and internal kinks of the boundary manifold, and they ensure that the expressions are gauge invariant and finite. Atiyah, Patodi, and Singer constructed general boundary terms which extend the gravitational index theorem for manifolds with boundaries \cite{AtiyahPatodiSinger1}\cite{AtiyahPatodiSinger2}. When the extra terms are taken into account, it can be shown that the Euler characteristic and second Chern class also vanish on our set of solutions. 

Heuristically this null result can be understood as follows. The index theorem gives a set of characteristic classes whose value is interpreted as the instanton number of the field configuration. The prototypical instanton in a manifold $\mathbb{R}\times \Sigma$ consists of an an initial data set at past infinity that evolves into an initial data set at future infinity that differs from the original by a gauge transformation that is not deformable to the identity\footnote{The past and future data sets do not need to be gauge related in order to have non-zero instanton number. However, for the purposes of illustration it is useful to restrict attention to these types of instantons.}. The key property is that the space of connections admits {\it continuous paths} interpolating between the two data sets. Despite the past and future data sets being gauge equivalent these paths are not gauge orbits, since no {\it continuous} gauge transformation can connect the two configurations. Thus, it is the continuous path in the space of connections on $\Sigma$, interpreted as a history on $\mathbb{R}\times \Sigma$ that is the instanton with non-zero winding number. Thus, one can take two spatial slices of two different solutions $\ou{m_1}{A}{n_1}$ and $\ou{m_2}{A}{n_2}$ to be the bounding data sets, and easily construct a smooth connection interpolating between these two sets interpreted as an instanton configuration. In a follow-up paper \cite{RandonoMercuri}, these types of configurations are constructed, and their implications for quantum gravity are explored. However, the solutions $\ou{m}{A}{n}$ themselves do not represent {\it transitions} between two homotopically inequivalent sectors in the space of connections. Thus, they each have zero instanton number, explaining why the Nieh-Yan, Chern-Pontryagin, and Euler classes are all zero.  

Thus, we must seek a different approach to physically distinguish the solutions. The strategy we will employ here is to first identify a distinguished set of hypersurfaces by appealing to a set of isometries common to all the solutions. In particular, we will first identify a $\mathbb{Z}_2$ isometry of all the states, and identify a distinguished hypersurface as the identity of the $\mathbb{Z}_2$ symmetry. The purpose of this construction is that these hypersurfaces are distinguished in a diffeomorphism invariant fashion, since the construction only employs the available isometries, which can be identified in a diffeomorphism invariant fashion. Next we will construct an appropriate generalization of the 3--volume of the distinguished hypersurfaces for each solution, which is invariant under $Spin(3,1)_M\rtimes Diff_4(M)$. As we will see, the value of this quantity is directly related to $q$, and therefore serves to differentiate the solutions as physically distinct. 

We first note that in addition to the set of ten continuous Killing isometries, de Sitter space has a discrete isometries. Here we will use the term isometry in a generalized sense: A transformation $\{e,\omega\}\rightarrow \{e_{*}, \omega_*\}$ is said to be an isometry if $e_*=e$ and $\omega_*=\omega$. de Sitter space has the special property that it is time-reversal symmetric. More specifically, there exists a parameterized foliation of the manifold $M$ into spatially compact hypersurfaces $\Sigma_t\simeq \mathbb{S}^3$ with $t\in\mathbb{R}$, such that under the diffeomorphism $\Sigma_t\rightarrow \Sigma_{-t}$, together with an appropriate discrete transformation of the $SO(3,1)$ vector space, the resulting transformation is an isometry. The action of time reversal on the internal space is well-known from quantum field theory as the antiunitary T-transformation of CPT-symmetry. Acting on the Clifford algebra, the action of T is given by $T(\gamma^0)=-\gamma^{0}$, $T(\gamma^i)=\gamma^{i}$, $T(\gamma^{[i}\gamma^{j]})=\gamma^{[i}\gamma^{j]}$, and $T(\gamma^{[i}\gamma^{0]})=-\gamma^{[i}\gamma^{0]}$. The resulting transformation obtain by applying $T$ and the diffeomorphism $t\rightarrow-t$ is an isometry\footnote{It is worth explaining the potentially confusing behavior of the extrinsic curvature $K^i\equiv {\omega^i}_0=\frac{1}{\ell}\tanh(t/\ell)\, e^i$. Under $t\rightarrow -t$ clearly $K^i\rightarrow -K^i$. However, under the time reversal operation of the internal indices, we also have $T(K^i)=T({\omega^i}_0)=-K^i$. Thus, in total $K^i_*=K^i$. Since the extrinsic curvature occurs explicitly in the expression for $\ou{m}{e}{n}$, some care must be taken to ensure that the isometry is evaluated properly.}. Furthermore, since each of the new solutions are obtained from the de Sitter solution by applying a $\ou{m}{g}{n}\in Spin(4,1)_M$ that is invariant under time reversal ($\ou{m}{g}{n}{}_*=\ou{m}{g}{n}$), each of the new solutions inherits the $\mathbb{Z}_2$ isometry of de Sitter space. We now identify a distinguished hypersurface $\Sigma_0$ as the hypersurface that is stabilized under the $\mathbb{Z}_2$ isometry. In the ordinary de Sitter case, due to the presence of the Killing isometries, the foliation $\Sigma_t$ and the distinguished hypersurface $\Sigma_0$ are not unique: the hypersurface will be invariant under three spatial rotations and three spatial transvections, but it will not be invariant under boosts and hypersurface orthogonal transvections. However, since these transformations are themselves isometries, the geometric content is preserved, and in particular, the $Spin(3,1)_M\rtimes Diff_4(M)$ invariant observable we will construct is independent of the choice of two different hypersurfaces, $\Sigma_0$ and $\Sigma'_0$ that are related by a Killing isometry. Although we will not discuss the Killing isometries of the new solutions, by the same argument, the existence of such isometries and the resulting ambiguity of the distinguished hypersurface will not affect the physical observable.

Given such a foliation $\Sigma_t$ of de Sitter space, we define the vector $n^I$ to be the lift of the hypersurface-normal vector $\bar{n}$ to the $SO(3,1)$ vector space so that $n^I=\ou{0}{e}{0}{}^I(\bar{n})$. The vector field can then be extended to each of the geometries $\{\ou{m}{e}{n},\ou{m}{\omega}{n}\}$ by applying a gauge transformation to define $\ou{m}{n}{n}{}^I=\ou{m}{g}{n}{}^I_J\,n^J$. In the standard polar chart of de Sitter space, the parameter $t$ of $\Sigma_t$ can be identified with the time variable, and in the diagonal gauge, the vector is given by $n^I=(1,0,0,0)$. This vector is preserved by $\ou{m}{g}{n}$, thus we will simply write $n^I=\ou{m}{n}{n}{}^I$ in this gauge. At all points where the metric is non-degenerate, one can define a hypersurface normal, $\bar{n}=\frac{1}{\cos(q\chi)}\frac{\partial}{\partial t}$, and at these points $n^I =\ou{m}{e}{n}{}^I(\bar{n})$. However, whereas the normal vector in the tangent space, $\bar{n}$ is undefined when $\cos(q\chi)=0$, the internal vector $n^I$ is defined everywhere. Thus, we can now define the natural generalization of the three-volume of a hypersurface
\beq
{}^3\ou{m}{V}{n}(\Sigma)\equiv \frac{4}{3!}\int_\Sigma \star \,n\,\ou{m}{e}{n}\w \ou{m}{e}{n}\w \ou{m}{e}{n} =\int_{\Sigma} \frac{1}{3!}\epsilon_{IJKL}\,n^I\,\ou{m}{e}{n}{}^J\w\ou{m}{e}{n}{}^K\w\ou{m}{e}{n}{}^L\,.
\eeq
The invariant observable will be the spatial 3-volume of the distinguished hypersurface: ${}^3\ou{m}{V}{n}(\Sigma_0)$.

The expression is gauge invariant under local $SO(3,1)_M$ transformations, and since the hypersurface $\Sigma_0$ is distinguished in a diffeomorphism invariant manner (up to isometry transformations, which don't change the value of the integral) by exploiting the available $\mathbb{Z}_2$ symmetry, the observable charge is invariant under diffeomorphisms as well.

\subsection{Calculating the charge}
As we have argued, if the volume ${}^3\ou{m}{V}{n}(\Sigma_0)$ is different for the geometries $\{\ou{m}{e}{n},\ou{m}{\omega}{n}\}$, the geometries can be viewed as physically distinct with respect to the gauge symmetries of Einstein-Cartan gravity. The task is then to calculate the charge. In fact the invariant can be calculated analytically as we will now show.

We first note, that in the given gauge, the pull-back of the connection, $\ou{m}{A}{n}$ to $\Sigma_0$ under the inclusion map $\rho:\Sigma_0\rightarrow M$ is pure diagonal:
\beq
\rho_{*}\ou{m}{A}{n}=\left[\begin{matrix} (\ou{m}{w}{n}{}^i+\frac{1}{\ell}\ou{m}{E}{n}{}^i)\tau_i & 0 \\ 0 & (\ou{m}{w}{n}{}^j- \frac{1}{\ell}\ou{m}{E}{n}{}^j)\tau_j \end{matrix} \right] 
\eeq
where $\ou{m}{E}{n}{}^i\equiv \rho_* \ou{m}{e}{n}{}^i$ can be interpreted as a spatial triad on the three sphere, and  $\ou{m}{w}{n}{}^i\equiv \frac{1}{2}\epsilon^i{}_{jk}\,\rho_* \ou{m}{\omega}{n}{}^{jk}$ can be interpreted as an $SU(2)$-connection compatible with the triad. These new variables satisfy (denoting $w\equiv w^i \,\tau_i$ and $E\equiv E^j \, \tau_j$)
\beqa
D_{\ou{m}{w}{n}}\ou{m}{E}{n}&=& 0 \nn\\
R_{\ou{m}{w}{n}}=d\ou{m}{w}{n}+\ou{m}{w}{n}\w \ou{m}{w}{n} &=&- \frac{1}{\ell^2}\ou{m}{E}{n}\w\ou{m}{E}{n} 
\eeqa
defining the constant curvature, zero-torsion geometry on the three sphere $\Sigma_0$. The connection $\rho_*\ou{m}{A}{n}$ can be interpreted as a flat $Spin(4)$-connection on the 3-sphere. This Lie group can be split into two copies of $SU(2)$ using the isomorphism $Spin(4)\simeq SU(2)_\ua \times SU(2)_\da$. Correspondingly, the connection can be split into two $SU(2)$-connections:
\beq
\ou{m}{W}{n}\!{}_+\equiv \left[\begin{matrix} (\ou{m}{w}{n}{}^i+\frac{1}{\ell}\ou{m}{E}{n}{}^i)\tau_i & 0 \\ 0 & 0 \end{matrix} \right]  \quad \quad \ou{m}{W}{n}\!{}_-\equiv \left[\begin{matrix} 0& 0 \\ 0 & (\ou{m}{w}{n}{}^i-\frac{1}{\ell}\ou{m}{E}{n}{}^i)\tau_i  \end{matrix} \right]\,.
\eeq
Consider now the Chern-Simons invariants of the two flat connections. Since $\ou{m}{W}{n}\!{}_+=g^m_\ua \ou{0}{W}{0}\!{}_+ g^{-m}_{\ua}-(dg^m_\ua)g^{-m}_\ua$ and $\ou{m}{W}{n}\!{}_-=g^n_\da \ou{0}{W}{0}\!{}_- g^{-n}_{\da}-(dg^n_\da)g^{-n}_\da$ we have:
\beq
Y_{CS}[\ou{m}{W}{n}\!{}_+]=Y_{CS}[\ou{0}{W}{0}\!{}_+]+m \quad \quad Y_{CS}[\ou{m}{W}{n}\!{}_-]=Y_{CS}[\ou{0}{W}{0}\!{}_-]+n \,.
\eeq
Thus, we have
\beqa
\left(Y_{CS}[\ou{m}{W}{n}\!{}_+]-Y_{CS}[\ou{m}{W}{n}\!{}_-]\right) -\left(Y_{CS}[\ou{0}{W}{0}\!{}_+] -Y_{CS}[\ou{0}{W}{0}\!{}_-] \right) =m-n\,.
\eeqa
On the other hand, the difference between the two Chern-Simons functionals takes the simple form:
\beqa
Y_{CS}[\ou{m}{W}{n}\!{}_+]-Y_{CS}[\ou{m}{W}{n}\!{}_-]&=& \frac{2}{8\pi^2}\left(\int_{\Sigma_0}\frac{2}{\ell}\ou{m}{E}{n}\w R_{\ou{m}{w}{n}} +\frac{2}{3\ell^3}\ou{m}{E}{n}\w\ou{m}{E}{n}\w\ou{m}{E}{n} \right)\nn\\
&=& -\frac{1}{3\pi^2\ell^3}\int_{\Sigma_0}\ou{m}{E}{n}\w\ou{m}{E}{n}\w\ou{m}{E}{n}\nn\\
&=&-\ou{m}{V}{n}(\Sigma_0)\big/ 2\pi^2 \ell^3\,.
\eeqa
 Thus, in total we have
 \beq
 \ou{m}{V}{n}(\Sigma_0)=2\pi^2\ell^3\left(1-q \right)
 \eeq
 from which we can define the invariant charge of the generalized de Sitter spaces
 \beq
\ou{m}{\mathcal{Q}}{n}\equiv 1-\ou{m}{V}{n}(\Sigma_0)\big/ 2\pi^2\ell^3\,. 
 \eeq
 
Now, suppose we have two solutions $\mathfrak{G}_1=\{\ou{m_1}{e}{n_1},\ou{m_1}{\omega}{n_1}\}$ and $\mathfrak{G}_2=\{\ou{m_2}{e}{n_2},\ou{m_2}{\omega}{n_2}\}$. When are the solutions physically distinct with respect to the gauge group $Spin(3,1)_M\rtimes Diff_4(M)$? From the previous discussion, the two geometries will be physically distinct if $\ou{m_1}{\mathcal{Q}}{n_1}\neq \ou{m_2}{\mathcal{Q}}{n_2}$. Moreover, if $\ou{m_1}{\mathcal{Q}}{n_1}= \ou{m_2}{\mathcal{Q}}{n_2}$, then $\ou{m_1}{A}{n_1}$ and $\ou{m_2}{A}{n_2}$ are related by a gauge transformation of the form $\ou{m}{g}{m}$, which by (\ref{mmElement}) is in $Spin(3,1)_M$. 
Thus, we conclude that $\mathfrak{G}_1$ and $\mathfrak{G}_2$ are physically distinct solutions to the first order Einstein-Cartan field equations {\it if and only if} $\ou{m_1}{\mathcal{Q}}{n_1}\neq \ou{m_2}{\mathcal{Q}}{n_2}$. 
 
\section{Concluding Remarks}
There are strong indications that the current incarnation gravity may be the symmetry broken phase of a more fundamental gauge theory based on the gauge group $Spin(4,1)$. Without knowledge of a full theory wherein the local symmetry is dynamically broken to $Spin(3,1)$, we have demonstrated that there are extremely generic properties of the symmetry-reduced theory that can be viewed as relic features of the full theory. They are generic in the sense that they do not depend on the details of the full theory in question apart from its symmetry group. In particular, by exploiting this symmetry we have constructed an infinite class of solutions to the first order Einstein-Cartan field equations with a positive cosmological constant. The solutions are physically distinct solutions with respect to the symmetries of Einstein-Cartan gravity, but they have some new and interesting properties not apparent in more conventional solutions to the field equations.

From the perspective of ordinary general relativity based on a smooth non-degenerate metric, it may be tempting to disregard these solutions as unphysical. However, from the perspective of gravity as a gauge theory, these solutions are perfectly natural. Moreover, if the full theory retained exact, local $Spin(4,1)$ invariance, as opposed to simply invariance under the $Spin(3,1)$ subgroup, these solutions would be gauge related, and therefore physically equivalent. Thus, metric gravity (barring a loose interpretation of metric theories allowing for degenerate metrics) would not allow for such states whereas gauge gravity might. For these reasons, we suggest that rather than dismissing the solutions as unphysical, the existence of such solutions should be regarded as a distinguishing characteristic of gravity as a gauge theory versus gravity as a metric theory, and the former should be more thoroughly explored.

\section*{Acknowledgments}
I would like to thank Takeshi Fukuyama for discussions and collaboration on a project spurring this investigation, as well as Laurent Freidel, Lee Smolin, Malcolm Perry, and Leonard Susskind for lively discussions at the Perimeter Institute. I would also like to thank Steve Carlip, especially for pointing out that the metric from this construction need not be invertible. This research was supported in part by NSF grant OISE0853116, NSF grant PHY0854743, The George A. and Margaret M. Downsbrough Endowment and the Eberly research funds of Penn State.

\appendix
\section{Notation and Conventions \label{App}}
Throughout this paper we will work with four-dimensional Riemannian manifolds with metric signature $(-,+,+,+)$. In these conventions, the de Sitter and anti-de Sitter metrics are the {\it non-degenerate} configurations $\{M,e,\omega\}$ satisfying
\beqa
d\omega^{IJ}+\omega^I{}_K\w \omega^{KJ}=\pm\frac{|\Lambda|}{3}e^I\w e^J \quad \quad de^I+\omega^I{}_K\w e^K=0
\eeqa
where for de Sitter space the sign is ``$+$" and the topology is $M=\mathbb{R}\times \mathbb{S}^3$, and for anti-de Sitter the sign is ``$-$" and the topology is $M=\mathbb{R}^4$ (the universal cover of $\mathbb{S}^1\times \mathbb{R}^3$).
For the local de Sitter and anti-de Sitter Lie algebras, it is convenient to work in a Clifford algebra representation. The Clifford algebra is defined by the condition
\beq
\gamma^I \gamma^J+\gamma^J \gamma^I=2\,\eta^{IJ}
\eeq
with $\eta^{IJ}=diag(-1,1,1,1)$. For generality, we will work with the double covers of the de Sitter and anti-de Sitter groups, $SO(4,1)$ and $SO(3,2)$, which are respectively, $Spin(4,1)$ and $Spin(3,2)$. We will employ a generic, standard complex $4\times 4$ matrix representation of the algebra, specifying the Dirac representation when necessary. In this representation, the fundamental representation of $\mathfrak{spin}(3,1)$ is the span of the Dirac bilinears, $\frac{1}{2}\gamma^{[I}\gamma^{J]}$, and the de Sitter and anti-de Sitter algebras are respectively
\beqa
\mathfrak{spin}(4,1)= Span\{\ts{\frac{1}{2}}\gamma^{[I}\gamma^{J]} \,,\, \ts{\frac{1}{2}}\gamma_5\gamma^K \} \quad \quad \mathfrak{spin}(3,2)=Span \{ \ts{\frac{1}{2}}\gamma^{[I}\gamma^{J]} \,,\, \ts{\frac{1}{2}}\gamma^K \}\,.
\eeqa
The tetrad is naturally valued in the vector elements of the Clifford algebra, $e\equiv \frac{1}{2}\gamma_I\,e^I$, with normalization defined so that $Tr(e\otimes e)=\mathfrak{g}$, where $\mathfrak{g}$ is the metric tensor. The spin connection is valued in the Lorentz subalgebra, $\omega=\frac{1}{4}\gamma_{[I}\gamma_{J]}\,\omega^{IJ}$, and its curvature is $R[\omega]\equiv \frac{1}{4}\gamma_{[I}\gamma_{J]}\,R^{IJ}=d\omega+\omega\w\omega$. The internal $Spin(3,1)$ dual is given by $\star\equiv -i\gamma_5=\gamma^0\gamma^1\gamma^2\gamma^3=\frac{1}{4!}\epsilon_{IJKL} \gamma^I\gamma^J\gamma^K\gamma^L$, with $\epsilon_{0123}=-\epsilon^{0123}=1$. Under an integral, the trace over the Clifford algebra is assumed so that, for example
\beqa
\int_{M}\star R\w R=\frac{1}{4}\int_{M}\epsilon_{IJKL}R^{IJ}\w R^{KL}\,.
\eeqa

\bibliography{MasterBibDesk}

\providecommand{\href}[2]{#2}\begingroup\raggedright\begin{thebibliography}{10}

\bibitem{Utiyama:1956sy}
R.~Utiyama, ``{Invariant theoretical interpretation of interaction},'' {\em
  Phys. Rev.} {\bf 101} (1956)
1597--1607.

\bibitem{Kibble:1961ba}
T.~W.~B. Kibble, ``{Lorentz invariance and the gravitational field},'' {\em J.
  Math. Phys.} {\bf 2} (1961)
212--221.

\bibitem{MMoriginal}
S.~Macdowell and F.~Mansouri, ``Unified geometric theory of gravity and
  supergravity,'' {\em Physical Review Letters} {\bf 38} (April, 1977)
  739--742.

\bibitem{West:1978Lagrangian}
P.~C. West, ``{A Geometric Gravity {L}agrangian},'' {\em Phys. Lett.} {\bf B76}
  (1978)
569--570.

\bibitem{Stelle:1979aj}
K.~S. Stelle and P.~C. West, ``{Spontaneously Broken {d}e {S}itter Symmetry and
  the Gravitational Holonomy Group},'' {\em Phys. Rev.} {\bf D21} (1980)
1466.

\bibitem{Stelle:1979va}
K.~S. Stelle and P.~C. West, ``{d}e {S}itter gauge invariance and the geometry
  of the {E}instein-{C}artan theory,'' {\em J. Phys.} {\bf A12} (1979)
L205--L210.

\bibitem{Fukuyama:Hamiltonian}
T.~Fukuyama and K.~Kamimura, ``Hamiltonian formulation of gauge theory of
  gravitation: Pure gravity case,'' {\em Nuovo Cimento} {\bf B74} (1983) 93.

\bibitem{Fukuyama:1984}
T.~Fukuyama, ``de {S}itter invariant gravity coupled with matters and its
  cosmological consequences,'' {\em Annals of Physics} {\bf 157} (1984) 321.

\bibitem{Fukuyama-Ikeda}
N.~Ikeda and T.~Fukuyama, ``Fermions in (anti) de {S}itter gravity in four
  dimensions,'' \href{http://www.arXiv.org/abs/arXiv:0904.1936}{{\tt
  arXiv:0904.1936}}.

\bibitem{Starodubtsev:MMgravity}
A.~Starodubtsev and L.~Freidel, ``Quantum gravity in terms of topological
  observables,'' \href{http://www.arXiv.org/abs/arXiv:hep-th/0501191}{{\tt
  arXiv:hep-th/0501191}}.

\bibitem{Bengtsson:1999ia}
I.~Bengtsson and S.~Holst, ``{de Sitter space and spatial topology},'' {\em
  Class. Quant. Grav.} {\bf 16} (1999) 3735--3748,
\href{http://www.arXiv.org/abs/gr-qc/9906040}{{\tt gr-qc/9906040}}.

\bibitem{WittMorrowJones:1993}
J.~Morrow-Jones and D.~M. Witt, ``{Inflationary initial data for generic
  spatial topology},'' {\em Phys. Rev.} {\bf D48} (1993)
2516--2528.

\bibitem{Scannell:Thesis}
K.~Scannell, {\em {Flat conformal structures and the classification of de
  Sitter manifolds}}.
\newblock PhD thesis, University of California, Los Angeles, 1996, 1999.

\bibitem{Witt:Designer}
K.~Schleich and D.~M. Witt, ``{Designer de Sitter Spacetimes},''
\href{http://www.arXiv.org/abs/0807.4559}{{\tt 0807.4559}}.

\bibitem{Carlip:Book}
S.~Carlip, {\em Quantum Gravity in 2+1 Dimensions}.
\newblock Cambridge University Press, December, 2003.

\bibitem{Matschull:1995nf}
H.-J. Matschull, ``{Three-dimensional canonical quantum gravity},'' {\em Class.
  Quant. Grav.} {\bf 12} (1995) 2621--2704,
\href{http://www.arXiv.org/abs/gr-qc/9506069}{{\tt gr-qc/9506069}}.

\bibitem{Baadhio:1992jn}
R.~A. Baadhio, ``{Knot theory, exotic spheres and global gravitational
  anomalies},''. In *Dayton 1992, Proceedings, Quantum topology* 78-90.

\bibitem{Zanelli:NiehYan}
O.~Chandia and J.~Zanelli, ``Torsional invariants, instantons and chiral
  anomaly on spaces with torsion,'' {\em Phys. Rev. D} {\bf 55} (1997)
  7580--7585, \href{http://www.arXiv.org/abs/arXiv:hep-th/9702025}{{\tt
  arXiv:hep-th/9702025}}.

\bibitem{HofmannMorris:Groups}
K.~H. Hofmann and S.~A. Morris, {\em The Structure of Compact Groups: a primer
  for students, a handbook for the expert}.
\newblock {d}e {G}ruyter Studies in Mathematics. Walter de Gruyter, 2nd~ed.,
  2006.

\bibitem{Instantons:Examples}
J.~Szczesny, M.~Biesiada, and M.~Szydlowski, ``{Topological quantum numbers and
  curvature -- examples and applications},'' {\em Int. J. Geom. Meth. Mod.
  Phys.} {\bf 6} (2009) 533--553,
\href{http://www.arXiv.org/abs/arXiv:0810.2911}{{\tt arXiv:0810.2911}}.

\bibitem{Eguchi:1980jx}
T.~Eguchi, P.~B. Gilkey, and A.~J. Hanson, ``{Gravitation, Gauge Theories and
  Differential Geometry},'' {\em Phys. Rept.} {\bf 66} (1980)
213.

\bibitem{AtiyahPatodiSinger1}
M.~F. Atiyah, V.~K. Patodi, and I.~M. Singer, ``{Spectral asymmetry and
  Riemannian Geometry 1},'' {\em Math. Proc. Cambridge Phil. Soc.} {\bf 77}
  (1975)
43.

\bibitem{AtiyahPatodiSinger2}
M.~F. Atiyah, V.~K. Patodi, and I.~M. Singer, ``{Spectral asymmetry and
  Riemannian geometry 2},'' {\em Math. Proc. Cambridge Phil. Soc.} {\bf 78}
  (1976)
405.

\bibitem{RandonoMercuri}
S.~Mercuri and A.~Randono, ``{The Immirzi parameter as an instanton angle}.''
  In preparation, 2010.

\end{thebibliography}\endgroup

\end{document}